\documentclass{ws-ijfcs}
\usepackage{amsmath}
\usepackage{amssymb}
\usepackage{graphicx}
\usepackage{algorithm}
\usepackage{algorithmicx}
\usepackage{algpseudocode}
\usepackage{multicol}
\floatname{algorithm}{Algorithm}

\usepackage{enumerate}
\usepackage{url}
\begin{document}

\markboth{Authors' Names}
{Instructions for Typing Manuscripts $($Paper's Title\/$)$}

%
\catchline{}{}{}{}{}
%

\title{Strict Linearizability and Abstract Atomicity}

\author{Tangliu Wen}
\address{State Key Laboratory of Software Engineering, School of Computer Science,Wuhan University,
 Wuhan, 430072, China\\
 \email{wqtlglk@163.com}
}

\author{Lan Song}

\address{State Key Laboratory of Software Engineering, School of Computer Science,Wuhan University,
 Wuhan, 430072, China
}

\maketitle

\begin{history}
\received{(Day Month Year)}
\accepted{(Day Month Year)}
\comby{(xxxxxxxxxx)}
\end{history}

\begin{abstract}
Linearizability  is a commonly accepted consistency condition for concurrent objects.
Filipovi\'{c} et al. show that linearizability is equivalent to observational refinement.
However, linearizability  does not permit concurrent objects to share memory spaces with their client programs.
We show that linearizability (or observational refinement) can be broken even though a client program of an object accesses the shared memory spaces without interference from the methods of the object.
In this paper, we present strict linearizability which lifts this limitation and can ensure client-side traces and final-states equivalence even in a relaxed program model allowing clients to directly access the states of concurrent  objects. We also investigate several important properties of strict linearizability.

At a high level of abstraction, a concurrent object can be viewed as a concurrent implementation of an abstract data type (ADT).  We also present a correctness criterion for relating an ADT and its concurrent implementation, which is  the combination of linearizability and data abstraction and can ensure observational equivalence.   We also  investigate its relationship with strict linearizability.
\end{abstract}

\keywords{Concurrent objects; linearizability; data abstraction; observational equivalence; atomicity.}

\section{Introduction}
Correctness conditions for concurrent objects generally require that each concurrent execution of an object is equivalent to a legal sequential execution
of either the object or an ADT. Different correctness conditions are distinguished by their different interpretation of the term ``equivalence''. Linearizability [8], sequential consistency [29] and quiescent consistency [14] have been widely accepted consistency conditions for concurrent objects.

 Filipovi\'{c} et al. [17] show that linearizability is equivalent to observational refinement---for a linearizable object $Z$, its corresponding ADT $A$ and any client program $P$, every observable behavior of $P(Z)$ can also be observed by $P(A)$, where $P(Z)$ denotes the client program $P$ that uses the object $Z$. Thus, clients can replace the fine-grained $Z$ with the more abstract coarse-grained $A$ to simplify reasoning.

However, linearizability assumes a complete isolation between an object and its client programs, and does not permit them to run in shared memory spaces.
The example in Section 2 shows that linearizability (or observational refinement) can be broken even though a client program of an object accesses the shared memory spaces without interference from the methods of the object.
A key reason for this is that
linearizability cannot ensure that a concurrent execution of an object and its corresponding sequential execution have the same final states.
In some applications,
concurrent objects need to share memory spaces with their client programs and permit them  to access the shared spaces via  atomic memory read/write actions.
In this cases,  atomicity specifications of concurrent objects should capture the above state consistency.

For example, RDCSS is part of the implementation of multiple compare-and-swap (MCAS) [20,10].
In MCAS, memory spaces are accessed via atomic memory read/write actions or the methods of RDSS.
Thus,  RDCSS must ensure that its linearizability cannot be broken by the atomic memory read/write actions of MCAS.
Furthermore, MCAS needs to share memory spaces with their client programs and  permits them  to access the shared spaces.
As another example, consider the atomic classes from
the java.util.concurrent.atomic package, such as the AtomicInteger class, the AtomicBoolean class.
Client programs can access the atomic variables (i.e.
instances of the atomic classes) via the methods of the classes or  atomic read/write actions.

In this paper, we present  strict linearizability, a correctness criterion aimed at lifting the above limitation.
We also show the following several properties of strict linearizability which linearizability cannot capture.
\begin{enumerate}[$\bullet$]
\item Strict linearizability can ensure
client-side traces and final-states equivalence.
Informally, for a strict linearizable object $Z$, any client program $P$,
$P(Z)$ has the same client-side traces and final states as
$P(Ato\_Z)$  even in a program model allowing $P$ to directly access the  states of $Z$ in some compatible ways.
Here, $Ato\_Z$ denotes an atomic version of $Z$  which
 complies with a sequential specification   of $Z$ and
  can be obtained by using  atomic regions to protect each method of $Z$.
 \item Strict linearizability can provide a strong termination-preserving property. For example,
we show that for a strict linearizable and purely-blocking object, a  program  using
the object diverges iff the  program  using its atomic version diverges.
Thus, while proving termination of a program using such an object, it is sufficient to replace the object with its atomic version.
\item For a strict linearizable object, its sequential specification can serve as ``maximal" atomicity abstraction (Theorem 17)---for a strict linearizable object $Z$, in order to verify whether $Z$  is a concurrent implementation of an ADT $A$, it is sufficient to check whether its sequential specification satisfies the specification of  $A$.  Obviously, verifying the latter is easier than verifying the former.
\end{enumerate}

In this paper, we refer to a sequential specification of a concurrent object as its atomicity specification.
A concurrent  object  satisfies its atomicity specification iff it is strict linearizable.
Most concurrent objects we know of ensure strict linearizability. For example, even many  subtle concurrent  objects, such as  RDCSS, MCAS, the pair snapshot algorithm [9], the MS lock-free queue [18], the lazy list algorithm [19] are  strict linearizable.

At a high level of abstraction, a concurrent object can be viewed as a concurrent implementation of an ADT.
What does it mean for a concurrent object to be an implementation of an ADT? Like in the sequential setting, data abstraction in the concurrent setting should also ensure the important representation independence property.
We state the representation independence property in terms of observational equivalence---two correct implementations of an ADT are observationally indistinguishable by clients of the ADT.
Linearizability is not sufficient to capture the representation independence property because it only ensures  observational refinement, not observational equivalence.
Thus, new observable behaviours can be introduced when clients replace a linearizable object with its corresponding ADT to simplify reasoning about  their programs (see the example in Subsection 6.2).

In this paper, we propose a correctness criterion for relating an ADT and its  concurrent
implementation,  which combines linearizability and data abstraction, and can ensure  observational equivalence.
Thus, like in the sequential setting, clients do not need to know the implementation details and internal synchronization mechanisms of concurrent objects, and can use the ADTs interfaces to reason about their programs. We refer to such an ADT specification as an abstract atomicity specification of its corresponding concurrent object.
We also investigate the relationship between atomicity specification and abstract atomicity specification.
As is mentioned above, for a strict linearizable object, its sequential specification can serve as ``maximal" atomicity abstraction. We moreover show the proof
obligations which can help establish atomicity in terms of abstract atomicity.

\section{Motivating Example}

In this section, we show that linearizability cannot ensure that a concurrent execution of an object and its corresponding sequential execution have the same final states.
A key reason for this is that linearizability is a property of externally-observable behaviors (i.e. histories) of concurrent objects. Informally, a history  consists only of input arguments and return values of the called methods of concurrent objects, not the internal states of concurrent objects.
When there is a complete isolation between an object and its client programs,
the inconsistent states cannot be observed by the client programs.
However, linearizability (observational refinement) can be broken even though  client programs access the internal states of an object without interference from the methods of the object. In this case, clients draw false conclusions when they reason about their programs in terms of the sequential specification of the object.

\begin{multicols}{2}
\multicolsep=2pt plus 2pt minus 2pt
\noindent
\tiny
class Queue\{\\
int back:=1; \\
data\_t[] items;\\
void Enqueue(data\_t v);\\
data\_t Dequeue( );\\
\}\\
void Enqueue(data\_t v)\{\\
L0 local t;\\
L1 t:=INC(back);\\
L2 items[t]:=v;\\
\}\\
data\_t  Dequeue()\{\\
L3 local temp,range;\\
L4 while(true)\{\\
L5 \quad temp:=null;\\
L6 \quad range:=back-1;\\
L7  \quad for(int i:=1;i<=range;i++)\{\\
L8 \quad \quad temp:=swap(items[i],null)\\
L9 \quad \quad if (temp!=null)\\
L10 \quad \quad return temp; \}\\
\indent  \} \}
\end{multicols}
\hrule
\begin{center} \tiny \textbf{Fig. 1.\,} the HW queue \end{center}

Fig. 1  shows  the HW queue.
The queue is represented as an infinite size array, $items$, and an integer variable, $back$, holding the smallest index in the unused part of the array. The index of the array starts with 1, and the variable $back$ is initialized to 1. The algorithm assumes each element of the array is initialized to a special value $null$.
The HW queue is linearizable with respect to a specification of a standard queue data type [1].

Consider the following program $P(HW)$:
\[\rm{HW.Enqueue(\textnormal{`}c\textnormal{'})}\;\parallel \; \rm{HW.Enqueue(\textnormal{`}d\textnormal{'})}\;\parallel \; \rm{HW.Dequeue()}\]
The program $P(HW)$  has four possible final states shown in Fig. 2.
However, the program $P(Ato\_HW)$  has only two possible final states shown in Fig. 2(c) and Fig. 2(d).
$Ato\_HW$ denotes the atomic version of  the HW queue, which
 complies with the sequential specification of the HW queue (see Section 5).
\begin{figure}[ht]
 \centering
  \includegraphics[height=0.2\textwidth]{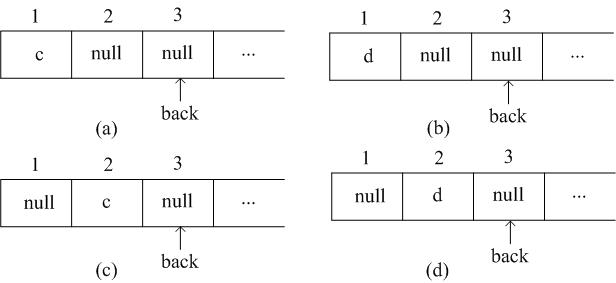}
\begin{center} \tiny \textbf{Fig. 2.}  Four possible final states of $P(HW)$  \end{center}
 \end{figure}

Obviously, $P(HW)$ and $P(Ato\_HW)$ have different final states.
The reason for this is that the final state of a concurrent execution may be inconsistent with that of the sequential execution whose history is a linearization of the history of the concurrent execution.
Consider an execution of $P(HW)$  generating the possible final state in Fig. 2(a), as shown in Fig. 3.
\begin{figure}[ht]
  \centering
  \includegraphics[height=0.15\textwidth]{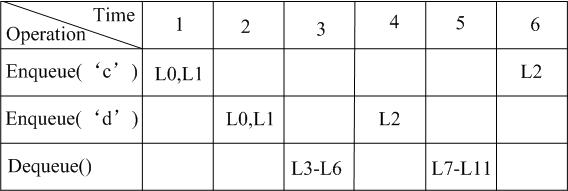}
\begin{center} \tiny \textbf{Fig. 3.} A concurrent execution of $P(HW)$  \end{center}
\end{figure}
By executing $INC$ command (line L1), the Enqueue(`c') operation reserves array position $1$ and the Enqueue(`d')  operation reserves array position $2$. The Enqueue(`d')  operation stores $\textnormal{`}d\textnormal{'}$ before the Enqueue(`c') operation stores $\textnormal{`}c\textnormal{'}$.
The Dequeue operation starts to traverse the array after the Enqueue(`d')  operation stores $\textnormal{`}d\textnormal{'}$  and returns before the Enqueue(`c') operation stores $\textnormal{`}c\textnormal{'}$. Thus the final state of the execution is that  $items[1]$ is $\textnormal{`}c\textnormal{'}$ and other elements of the array are $null$.
The only sequential execution which
produces a linearization of the history of the concurrent execution depicted above is:
 \[\rm{HW.Enqueue(\textnormal{`}d\textnormal{'})};\rm{HW.Enqueue(\textnormal{`}c\textnormal{'})};\rm{HW.Dequeue()}\]
 The final state of the sequential  execution is that $items[2]$ is $\textnormal{`}c\textnormal{'}$, other elements of the array are $null$.
Thus,
 the inconsistent state can be observed by client programs
 even though the client programs  access the elements of the array $items$  without interference from the methods of the HW queue.
 In this case, clients  draw  false conclusions when they reason about their programs  in terms of the sequential specification of the HW queue.

\section{Trace Model}
\subsection{Characterizing Behaviours of Concurrent objects via Trace Model}

In the concurrent setting, a concurrent object  provides a set of methods, which can be called  concurrently by its client programs.
For a concurrent object  $Z$, let $Zop$ denote a set of the methods of  $Z$. Let $P(Z)$ denote a client program $P$ that uses the object $Z$.
 For simplicity, we assume  each method takes one argument and returns a value using the command $ret (E)$.
 The methods are defined by declarations of the form $f(x):C;ret (E)$.
Here $f$ is a method name,  $x$ is a formal argument, $C;ret(E)$ is a method body.
The method calls  are of the form $x:=Z.f(E)$.

\begin{eqnarray*}
&&{\rm E} :: = {\rm n} {\kern 2pt}| {\kern 2pt}  {\rm x}{\kern 2pt} | {\kern 2pt}  {{\rm E} } + {{\rm E} }{\kern 2pt}|\ldots\\
&&{\rm B} :: = {\rm true} {\kern 2pt}|{\kern 2pt} {\rm false}{\kern 2pt} |{\kern 2pt} {\rm E} = {\rm E}{\kern 2pt}|
{\kern 2pt} {\rm E} \le {\rm E}{\kern 2pt}|\ldots
\\
&&{\rm I} :: = {\rm x} : = [{\rm E}]{\kern 2pt}
|[{\rm E} ]:={\rm E}{\kern 2pt}|{\kern 2pt}  {\rm x} \!: = {\rm E} {\kern 2pt}
|{\kern 2pt}{\rm x}: = {\rm cons}({\rm E} ){\kern 2pt}|{\kern 2pt}\ldots
\\
&&{\rm C} :: = {\rm I} {\kern 2pt}|{\kern 2pt}{\rm x}\! : = {\rm Z.f(E)}
{\kern 2pt} |{\kern 2pt}{\rm C};{\rm C}{\kern 2pt}|{\kern 2pt}{\rm if}  {\kern 2pt} {\rm B}  {\kern 2pt}  {\rm then} {\kern 3pt} {\rm C} {\kern 3pt} {\rm else}  {\kern 3pt} {\rm C} {\kern 2pt} {\kern 2pt}|
 {\kern 2pt}{{\rm while}} {\kern 2pt} {\rm B}  {\kern 2pt} {{\rm do}} {\kern 2pt} {\rm C}{\kern 2pt}|{\kern 2pt} \langle{\rm C }\rangle\\
&& Zop:: =\{f_1(x_1):C_1;ret (E_1),\ldots, f_n(x_n):C_n;ret (E_n)\}
\\
&&{\rm P(Z)}:: ={\rm C}\!\parallel \!\cdots\!\parallel \!{\rm C}
\end{eqnarray*}
\begin{center}\begin{footnotesize} \textbf{Fig. 4.} \;\; Syntax of the Programming Language\end{footnotesize}\end{center}

  $P(Z)$ contains several sequential commands, each of which is executed by a thread,  as shown in Fig. 4. $I$ is a set of primitive instructions, $ cons$ is  an allocating  memory cells  command, $x $$:= $$[E]$  and $ [E] $$:=$$ E$  are  reading  and writing memory cells commands respectively. $\langle{\rm C }\rangle $ is an atomic region. An atomic action is either a primitive instruction or an atomic region. Let $A$ be a set of atomic actions.

Let $M$ be a set of method names, $T$
 be a set of thread identifiers. We refer to a method call as an operation.
Let $O$ be a set of operation identifiers which is used to identify every method call.
 An action label is either an
invocation of an operation, a response by an invoked operation, or an atomic action.
 An  event consists of  a thread identifier and an action label and an operation identifier (if an event of an object), and can be one of the following forms:
\[Event ::=(t, inv(m,v),o)\; |\;(t, a,o)\; |\;(t, ret(v),o)\; |\;(t,  a)\]
where $t \in T$, $o \in O$, $m\in M$, $a \in A$,  $v\in Values$.
The event $ (t, inv(m,v),o)$ represents  an invocation event of a method $m $  with an argument value $v$ which is performed by a thread $t$ and is identified by an operation identifier $o$.
$(t,a,o)$  represents an event  of an operation $o's$  body.
 $(t, ret(v),o) $ represents a response  event of an operation $o$   with a return value $v$.
  $(t, a)$ represents a client program's event which is
 performed by a thread $t$.
For an event  $e$,  Let $ Thr(e)$  denote  its thread identifier,  $Lab(e)$  denote its  label,  $Op(e)$ denote its operation identifier.
We sometimes omit the first and third fields of events when they are irrelevant to discussions.
Let $invAct$ be a set of invocation events,
$resAct$ be a set of  response  events.
An  invocation  event $e_1\in invAct $ matches a response  event $ e_2 \in resAct$,
denoted by $ e_1 \rightsquigarrow e_2$, if $Op(e_1)=Op(e_2)$.

\begin{align*}
\tiny
&[\kern-0.15em[ E ]\kern-0.15em]_t\subseteq Tr\times Values
\\
&[\kern-0.15em[ a ]\kern-0.15em]_t=\begin{cases}
(a,t),\quad\rm{if\,} a\in A \,\rm{ occurs\,in \,a\, cilent \, program; }
\\ (a,o,t),\quad\rm{if \,} a\in A \,\rm{ occurs\,in \, an \,operation \,o.}
\end{cases}
\\
& [\kern-0.15em[ ret(E) ]\kern-0.15em]_t=\{\rho^\smallfrown(t,o,ret(v))\; | \; (\rho, v) \in [\kern-0.15em[E]\kern-0.15em]_t\}
\\
&[\kern-0.15em[C_1;C_2 ]\kern-0.15em]_t=[\kern-0.15em[C_1 ]\kern-0.15em]_t[\kern-0.15em[C_2 ]\kern-0.15em]_t=\{\rho_1^\smallfrown\rho_2\;|\;\rho_1\in[\kern-0.15em[C_1 ]\kern-0.15em]_t
\wedge\rho_2\in[\kern-0.15em[C_2 ]\kern-0.15em]_t\}
\\
&[\kern-0.15em[ if  {\kern 2pt}  B  {\kern2pt}  then {\kern 3pt} C_1 {\kern 3pt}  esle  {\kern 3pt} C_2 ]\kern-0.15em]_t=[\kern-0.15em[B]\kern-0.15em]_t^{true}[\kern-0.15em[C_1 ]\kern-0.15em]_t\;\cup\;[\kern-0.15em[B]\kern-0.15em]_t^{false}[\kern-0.15em[C_2 ]\kern-0.15em]_t
\\
&[\kern-0.15em[while {\kern 2pt}  B  {\kern 2pt}  do {\kern 3pt} C  ]\kern-0.15em]_t=([\kern-0.15em
[B]\kern-0.15em]_t^{true}[\kern-0.15em[C ]\kern-0.15em]_t)^*[\kern-0.15em[B]\kern-
0.15em]_t^{false}
\;\cup\;([\kern-0.15em[B]\kern-0.15em]_t^{true}[\kern-0.15em[C]\kern-
0.15em]_t)^\omega
\\
&[\kern-0.15em[  x=z.f(E)]\kern-0.15em]_t=\{ \rho_1^\smallfrown(t,o,inv(f,n))^\smallfrown\rho_2^\smallfrown
(t,o,ret(v))^\smallfrown(t,x:=v)\; | \; (\rho_1, n) \in [\kern-0.15em[E]\kern-0.15em]_t
\\
&\wedge\; \rho_2^\smallfrown(t,o,ret(v))\in  [\kern-0.15em[f_{body}]\kern-0.15em]_t\}
\\
&[\kern-0.15em[  C_1\!\parallel \! C_2 ]\kern-0.15em]=
\bigcup\{\lambda_1\;\vert\kern-0.4em\vert\kern-0.4em\vert
\;\lambda_2\;|\;
\lambda_1\in[\kern-0.15em[C_1]\kern-0.15em]_{t1}
\wedge\lambda_2\in[\kern-0.15em[C_2]\kern-0.15em]_{t2} \}
\end{align*}
\hrule
\begin{center}\begin{footnotesize} \textbf{Fig. 5.} \;\;  Trace Semantics of Commands and programs\end
{footnotesize}\end{center}

A trace is a sequence of  events. For a trace $\lambda$,
let $|\lambda|$  denote the length of the trace; $|\lambda|=\omega$ if $\lambda$ is infinite.
Let $<_a$  denote a happened-before order on events; for two events $c$ and $d$ in a trace,  $c<_a d$ if $c$ precedes $d$ in the trace.

 The semantics  of  commands and programs is defined in terms of traces [28], and it is shown in Fig. 5. Let $Tr $ be a set of all traces.
We write $ \rho_1^\smallfrown\rho_2$ for the trace obtained by concatenating $\rho_1$ and $ \rho_2$; when
$\rho_1$ is infinite this is just $ \rho_1$.
Let $[\kern-0.15em[ C ]\kern-0.15em]_t$
be a set  of  traces of a sequential command $C$, which is parameterized by a thread (which runs the command).
$[\kern-0.15em[ E ]\kern-0.15em]_t$ is a set of all $(\rho,v)$ such that $E$ evaluates to $v$ along the trace $\rho$. $[\kern-0.15em[B]\kern-0.15em]_t^{true}$ is a set of all $\rho$ such that
$(\rho,true)\in [\kern-0.15em[B]\kern-0.15em]_t$.
 $f_{body}$ denotes the  body of the method $f$.
The operator $\vert\kern-0.4em\vert\kern-0.4em\vert$ forms a set of all possible interleavings of two sequences.

\subsection{Client-Side Traces and Final States}
We assume that states of  client programs are disjoint from ones of  concurrent objects.
The assumption is used in the standard  notion of linearizability [8].
For a program $P(Z)$, a valid state  is  $(\sigma_c, (\sigma_z, u))$, where $\sigma_c$
is a  state of the client program $P$, $(\sigma_z, u)$ is a configuration of the object $Z$.
 $\sigma_z$ is a state of  $Z$, which records the values of the concurrent object's shared data and pointer variables.
Let $ l \in Lop $ denote a local state of an operation.
 $u\in U$ represents a mapping $ u :O\rightarrow Lop $, which maps every operation
to their local states. Let $\phi \in U $ be an empty mapping; $ u=\phi $ when all operations do not begin to execute.

A transition is a triple of the form  $ \sigma \xrightarrow{e} \sigma'$, where $ \sigma $ and $ \sigma'$ are states and $ e$ is an event.
For example, a transition
$(\sigma_c,(\sigma_z,u))$ $\xrightarrow{e} (\sigma_c',(\sigma_z,u))$
 characterizes the effect that a state $ \sigma_c$
 can be transformed into a state $ \sigma_c'$  by an event $e$ of a client program.
We use  $abort$ to denote an error state.
A transition  $(\sigma_c,(\sigma_z,u))$ $\xrightarrow{e} abort $ characterizes the effect that an event $e$  leads to  a runtime error.

A terminating execution $\pi$ of a program $P(Z)$ from an initial state $(\sigma_{c0},$ $(\sigma_{z0},\phi))$ is a finite sequence of the form
$(\sigma_{c0},(\sigma_{z0},\phi))$ $\xrightarrow{e_1} (\sigma_{c1},(\sigma_{z1},u_1))
\xrightarrow{e_2},\cdots,\xrightarrow{e_n}(\sigma_{cn},(\sigma_{zn},u_n))$, where the trace $e_1e_2\cdots e_n\in [\kern-0.15em[  P(Z)]\kern-0.15em]$.

For an execution $\pi$ of $P(Z)$,
let $tr(\pi)$ denote the trace generated by the execution $\pi$,
let $tr(\pi)\lceil A_c$  denote the maximal subsequence of $tr(\pi)$ consisting of the  events of the client program $P$ (i.e., the projection of the trace $tr(\pi)$  to the  events of $P$),
let $tr(\pi))\lceil A_z$ denote the maximal subsequence of $tr(\pi)$ consisting of the events of  the object $Z$ (i.e., the projection of the trace $tr(\pi)$  to the  events of $Z$),
let $tr(\pi)\lceil t$  denote the maximal subsequence of $tr(\pi)$ consisting of the  events  performed by the thread $t$.

For a program $P(Z)$, let $(\sigma_{c},\sigma_{z})\xrightarrow{\lambda}{(\sigma_{c}',\sigma_{z}')}$ denote a terminating execution  of $P(Z)$ which starts from the state  $(\sigma_{c}, (\sigma_{z},\phi))$, terminates in the state $\sigma_{c}'$ of $P$ and the state $\sigma_{z}'$ of $Z$   and generates the finite trace $\lambda$; let $(\sigma_{c},\sigma_{z})\xrightarrow{\lambda}{^\omega\_}$  denote a divergent execution of $P(Z)$ which starts from the state  $(\sigma_{c},(\sigma_{z},\phi))$ and  generates the infinite trace $\lambda$ ;
let $(\sigma_{c},\sigma_{z})\xrightarrow{\lambda} { abort} $ denote
 an execution of $P(Z)$ which starts from the state $(\sigma_{c},(\sigma_{z},\phi))$, produces a runtime error and generates the finite trace $\lambda$ .

A divergent execution of $P(Z)$ could be caused by   divergences of  the client program $P$, divergences of $Z$   or a combination of both. Let $(\sigma_{c},\sigma_{z})\xrightarrow{\lambda}{^\omega_C\_}$  denote a divergent execution of $P(Z)$  caused by a divergence of the client program $P$ (i.e. $ |{\lambda}\lceil{A_c}|=\omega\wedge |{\lambda}\lceil{A_z}|\not =\omega$).

Given a program $P(Z)$, a state $\sigma_{c}$ of the client  program $P$ and a  state $\sigma_{z}$ of the object $Z$, the client-side traces of the program, denoted by $\mathcal{MT}[\kern-0.15em[ P(Z)(\sigma_{c},\sigma_{z}) ]\kern-0.15em]$,  and the final states of the program,   denoted by $\mathcal{MS}[\kern-0.15em[ P(Z)(\sigma_{c},\sigma_{z}) ]\kern-0.15em]$,  are defined as follows.

$\mathcal{MT}[\kern-0.15em[ P(Z)(\sigma_{c},\sigma_{z}) ]\kern-0.15em]=\{\lambda\lceil{A_c} {\kern 2pt} | {\kern 2pt} (\sigma_{c},\sigma_{z}) \xrightarrow{\lambda}{({\sigma_c}',{\sigma_z}')}
\vee {\kern 2pt}(\sigma_{c},\sigma_{z}) \xrightarrow{\lambda} {abort}{\kern 2pt}\vee {\kern 2pt}(\sigma_{c},\sigma_{z})
\xrightarrow{\lambda}{^\omega_C\_}
\}$

$\mathcal{MS}[\kern-0.15em[ P(Z)(\sigma_{c},\sigma_{z}) ]\kern-0.15em]=\begin{cases}
\{(\sigma_{c}',\sigma_{z}') \;|\; (\sigma_{c},\sigma_{z})\xrightarrow{\lambda}
{(\sigma_{c}',\sigma_{z}')}\}\\\;\cup\;
\{abort \;|\; (\sigma_{c},\sigma_{z})\xrightarrow{\lambda} {abort}\}
\\\;\cup\;\{\perp\;|\;(\sigma_{c},
\sigma_{z})\xrightarrow{\lambda}{^\omega_C\_}\}
\end{cases}$

\section{Strict Linearizability}

\subsection{Histories and Linearizability Relation. }
Linearizability is defined using the notion of histories. The history of an execution $\pi$, denoted by $H(\pi)$, is the maximal subsequence of $tr(\pi)$  comprised of the invocation and response events.

For a history $H$, let $ H(i) $ denote the $i^{th}$ element of $ H $.
A history is sequential if the event preceding each response event is its matching invocation event.
 A history $H$ is well-formed if for every thread $t$, $H\lceil t$ is sequential.
A history is complete if it is well-formed and every invocation event has a matching response event. An invocation event  is pending  if there is no matching response event to the invocation event.

We introduce the special response event for an aborted operation $o$, denoted by $(t,o,abort)$.
For an execution $\pi$,
let $resAbort(H(\pi))$ be the history gained  by adding  matching special response events for all aborted operations in $\pi $ to  the end of the history $ H(\pi) $.
For an incomplete history  $ H(\pi) $, a completion of $ H(\pi) $, is a complete history gained  by  adding some matching response events  to the end of $resAbort(H(\pi))$ and removing some pending invocation events within $resAbort(H(\pi))$.
Let $Compl(h)$ denote a set of all completions of the history $ h $.

 We use $ (\sigma_z,H,{\sigma_z}')$ to denote  a terminating execution of a concurrent object (i.e.  all invoked methods of the object  have returned in the execution), where  $\sigma_z $  is  the object's initial state, $ H$ is the history of the execution, $ {\sigma_z}' $ is   the object's  final  state.
Let $<_o$  denote  the happened-before order on operations; for two operations $o$ and $o'$, $o<_o o'$ if the response event of $o$ precedes the invocation event of $o'$.

\begin{definition} [Linearizability Relation {[}11{]}]
The linearization relation $\sqsubseteq$ on histories is defined as follows: $ H \sqsubseteq H' $ iff\\
1. $\forall t. H\lceil t = H'\lceil t$;\\
2. there exists a bijection $  \nu : \{1, \ldots, |H|\} \to \{1, \ldots, |H'|\}$ such
that $  \forall i.  H(i)= H'(\nu (i))$ and
$\forall i,j. i < j \wedge H(i) \in resAct \wedge H(j)\in invAct \Longrightarrow \nu(i) < \nu(j).$
\end{definition}

The first condition above requires that $H'$ is a permutation
of $H$; the second condition above requires that
the happened-before order between non-interleaved operations in  $H$  and  $H'$  is identical.
The following proposition shows that the linearizability relation is transitive.
\begin{proposition}
$H_1 \sqsubseteq H_2 \wedge  H_2 \sqsubseteq H_3 \Longrightarrow H_1 \sqsubseteq H_3$
\end{proposition}
The proof for the proposition appears in Appendix Section A.

\subsection{Strict Linearizability}
A sequential specification of an object is used to describe the semantics of the object in the absence of concurrency.

\begin{definition} [Sequential Specification]
For a concurrent object $Z$, let $ZState$ be a set of the  well-formed states of  $Z$,  $Zop$ be a set of the methods of  $Z$, $Input$  be a set of input values, $Output$ be a set of output values. A  sequential specification  of a method $op \in Zop$ is a partial function
$op: ZState\times Input\nrightarrow ZState\times Output$.
\end{definition}

For the sequential specifications of concurrent objects, only well-formed states need to be considered. For example,  a  state of  a  singly  linked list is well-formed only when there are  no loops or cycles in it.
Note that  the methods are defined as partial functions. A method  is total if it is defined in the set $ ZState\times Input$; it is partial if it is defined in a proper subset of the set $ ZState\times Input$.

For a sequential execution of a method $op$  starting from a state $ \sigma_z$  with an input $ in$, let $ (\sigma_z,in)op(\sigma_z',ret)$ denote that the execution is error-free, and terminates in a  state $ \sigma_z'$ with an output $ret$.

A  sequential execution of a method $op$
starting from a state $\sigma_z\in ZState $ with an input $ in\in Input $ is legal if  $op(\sigma_z,in)=({\sigma_z}',ret) \Longrightarrow (\sigma_z,in)op(\sigma_z',ret)$,
where
  $op(\sigma_z,in)=({\sigma_z}',ret)$  denotes that the result of applying the operation (or function) $op$ to an input $in$ and  a state $\sigma_z$ is the state ${\sigma_z}'$ and the return value $ret$.
A  sequential execution of an object is legal if the executions of all methods of the execution are legal.

\begin{definition}[Strict Linearizability]
A concurrent object $Z$ is  strict linearizable iff \\
1. for every execution $\pi$ of $Z$ starting from any well-formed state $\sigma_z$,  there exists a legal sequential execution $\pi'$ of $ Z$ starting from the state $\sigma_z$  and a history $h_c \in Compl(H(\pi))$  such that $ h_c\sqsubseteq H(\pi') $;\\
2. for every terminating execution $\pi: (\sigma_z,H_{con},{\sigma_z}')$ of $Z$, there exist a legal sequential execution $\pi':  (\sigma_z,H_{seq},{\sigma_z}') $ of $Z$  such that $H_{con}\sqsubseteq H_{seq}$.
\end{definition}

We refer to a sequential specification of a concurrent object as its atomicity specification.
A concurrent  object  satisfies its atomicity specification iff it is strict linearizable.
In order to distinguish between  strict linearizability
and  classical linearizability [8,24,25],
we call the latter general linearizability, which is formally defined as follows.

\begin{definition} [General Linearizability]
A concurrent object $Z$ is general linearizable w.r.t. an ADT $A$, if  for any execution $\pi$ of $Z$ starting from any well-formed state $\sigma_z$,
there exists a legal sequential execution $\pi'$ of $A$ starting from the state $AF(\sigma_z)$  and a history $h_c \in Compl(H(\pi))$  such that $ h_c\sqsubseteq H(\pi') $.
\end{definition}
Here $AF$ denotes an abstraction function which maps
the well-formed states of a concurrent object to the states of an ADT. The specifications for ADTs and abstraction functions will be explained in detail in Section 6.

\section{Properties of Strict Linearizability }

In this section, we show several important properties of strict linearizability which general linearizability does not  capture.

\subsection{Client-Side Traces and Final-States Equivalence }
For a concurrent object $Z$, let $Ato\_Z$ denote the atomic version of $Z$ in which every method  is protected by an atomic region.
We use the sequential specification of $Z$ to describe the semantic of $Ato\_Z$.
An operation $op$ of $Ato\_Z$ is executed atomically  if its precondition (i.e., the domain of  $op$) is true in  the current configuration,
otherwise it blocks execution from the current configuration.
If during a concurrent execution, its precondition becomes
true (similar to the spinlock, repeatedly testing the precondition), the operation $op$  can resume its execution.
When the precondition of an operation is true, the trace semantics of the operation is defined as follows:\\
$[\kern-0.15em[  x=Ato\_Z.\langle f(E)\rangle ]\kern-0.15em]_t=\{ \rho_1^\smallfrown\langle(t,o,inv(f,n)),\rho_2,(t,o,ret(v))
\rangle^\smallfrown(t,x:=v) \; | \; (\rho_1, n) \in [\kern-0.15em[E]\kern-0.15em]_t
\wedge \rho_2^\smallfrown(t,o,ret(v))\in  [\kern-0.15em[f_{body}]\kern-0.15em]_t\}$

Here, $ \langle(t,o,inv(f,n)),\rho_2,(t,o,ret(v))\rangle$ is
an atomic trace, i.e., it is interleaved with other events
as a single event.
The following theorem shows  that  strict linearizability can ensure client-side traces ans final-states equivalence. The proof for the theorem appears in Appendix Section B.

\begin{theorem}
 A concurrent object $Z$ is strict linearizable
iff for any client program $P$, any initial state $\sigma_{c}$ of $P$, any well-formed state $\sigma_{z}$ of  $Z$,\\
1. $ \mathcal{MT}[\kern-0.15em[ P(Z)(\sigma_{c},\sigma_{z}) ]\kern-0.15em]=\mathcal{MT}[\kern-0.15em[ P(Ato\_Z)(\sigma_{c},\sigma_{z}) ]\kern-0.15em] $,\\
2.
 $ \mathcal{MS}[\kern-0.15em[ P(Z)(\sigma_{c},\sigma_{z}) ]\kern-0.15em]=\mathcal{MS}[\kern-0.15em[ P(Ato\_Z)(\sigma_{c},\sigma_{z}) ]\kern-0.15em]$.
\end{theorem}
The first condition  shows that $P(Z)$ and $P(Ato\_Z)$ have the same client-side traces.
We call this property client-side traces equivalence.
This means that $P(Z)$ and $P(Ato\_Z)$ have the same
linear-time temporal properties of the client program.
 The second condition shows that  $P(Z)$ and $P(Ato\_Z)$ have the same final states.
Thus, for a strict linearizable concurrent object, clients do not need to know the internal synchronization details of the concurrent object and can design,  program and reason in terms of its sequential specification.

The theorem above is obtained in the program model where
there is a complete isolation between an object and its client programs.  We now relax the restriction of the program model by allowing client programs to share memory spaces with objects and to directly access the shared memory spaces via compatible atomic memory read/write actions.
Atomic memory read/write actions are said to be compatible if they satisfy either of  the following two restrictions:
\begin{enumerate}[$\bullet$]
\item The read/write actions do not interfere with the methods of concurrent objects, i.e., when  the write actions are executed, the methods of concurrent objects which are called before them  have finished; when  the read actions are executed, the methods which are called before them either have finished or do not modify the states of objects. The write actions maintain well-formed states of concurrent objects.
    \item If the  read/write actions are encapsulated into the methods of concurrent objects,
the ``new'' methods do not break strict linearizability (i.e., after adding the ``new'' methods, the concurrent objects are still strict linearizable).
\end{enumerate}
The following theorem shows  that  strict linearizability can provide the same guarantee even in the relaxed program model above.
The proof for the theorem appears in Appendix Section B.

\begin{theorem}
For a strict linearizable object $Z$ with a well-formed initial state, for any client program $P$,
$P(Z)$ and $P(Ato\_Z)$ have the same client-side traces and final states even in the relaxed program model above.
\end{theorem}

\subsection{Preservation of Termination }
In this subsection, we consider two  progress properties, minimal termination and purely-blocking progress [11].
 We show that for a strict linearizable object satisfying either of the two progress properties, a  program  using
the object diverges iff the program  using its atomic version diverges.
Thus, while proving termination of a program using such an object, we can soundly replace the object with its atomic version.

Note that the definition of  client-side traces (in Section 3) does not consider divergences of  concurrent objects. Thus, from client-side traces equivalence of Theorem 6, we get the following corollary.

\begin{corollary}
 For a strict linearizable object $Z$  with a well-formed initial state, any client program $P$, $P(Z)$ diverges by a divergence of the client program $P$  iff  $P(Ato\_Z)$ diverges by the same reason.
\end{corollary}

We now introduce a progress property called minimal termination.
An object satisfies  minimal termination iff for any client program $P$, every method of the object can terminate
if $P$ calls the methods finitely.
There are a variety of objects that satisfy minimal termination, e.g., wait-free, lock-free,
deadlock-free, starvation-free objects have such a progress guarantee.

For an object $Z$ satisfying minimal termination, if a client program $P$ calls its methods finitely, then $P(Z)$ cannot diverge by the divergences of $Z$.
Thus, in terms of Corollary 8, we get the following lemma.
\begin{lemma}
 For a strict linearizable object $Z$ satisfying minimal termination  with a well-formed initial state, any client program $P$,  $P(Z)$ diverges  iff  $P(Ato\_Z)$ diverges.
\end{lemma}

We now consider the purely-blocking progress [13], which is a weaker progress property than minimal termination. An object is purely-blocking [13] when at any reachable state, any pending method, if run in isolation will terminate or its entire execution does not modify states of the object.
Minimal termination restricts the methods of concurrent objects to be total. The purely-blocking progress permits  the methods of concurrent objects to be partial. For example, the $HW$ queue is purely-blocking [13] and its  $Dequeue()$ method is a partial method.
The following theorem shows that a purely-blocking and strict linearizable  object   possesses a strong termination-preserving property.
We also show that the theorem still holds for the relaxed program model in subsection 5.1. The proof for the theorem appears in Appendix Section C.

\begin{theorem}
For a strict linearizable and  purely-blocking  object $Z$  with a well-formed initial state, any client program $P$, $P(Z)$ diverges  iff  $P(Ato\_Z)$ diverges.
\end{theorem}
The HW queue cannot ensure the termination-preserving property, because it is not strict linearizable.  For example, consider the following program:
\begin{eqnarray*}
&&HW.Enqueue('c')\parallel HW.Enqueue('d') \parallel HW.Dequeue();\\
&&HW.item[1]=x; \;{\text{\small  /\kern-0.3em/ which are not interleaved with the called methods}} \\
&&HW.Dequeue()\parallel HW.Dequeue();
\end{eqnarray*}
The program above can diverge. However, when the program replaces  the HW  queue with its atomic version, the program  always terminates.

\section{ Abstract Atomicity   }

\subsection{Data Abstraction For Sequential Data Structures}

We use model-based specification [15] to define ADTs, where an ADT is considered as a set of abstract values together with a set of atomic methods; the methods are specified by defining how they affect the abstract values.

\begin{definition}[Abstract Data Type]
An ADT $ A $  is a tuple $(AState, \sigma_{a0}, $ $ Aop,Input,Output )$,
where $AState$ is a set of states; $\sigma_{a0}\in AState$ is the initial state;
$ Aop$ is a set of methods; $Input$ is a set of input values;
$ Output $ is a set of output values;
each method $op\in Aop $ is a mapping  $op : AState\times Input\nrightarrow AState\times Output$.
\end{definition}

Let $dom(op)$ denote the domain (i.e., precondition) of the method $op$. A method of $A$  blocks when it is called outside its domain.
In the sequential setting, for an ADT $A$ and its implementation (or representation) $Z$, abstraction function $AF: ZState\nrightarrow AState$ is used to map the well-formed states of $Z$ to the  states of $A$.
An abstraction function explains how internal structure of an implementation is viewed abstractly by clients of an ADT.
The function is surjective and thus every abstract state can be represented by one or more concrete states.
The renaming function $RF: Zop\rightarrow Aop$ is used to map the method names of $Z$ to the method names of $A$. The inverse of the function $RF$ is denoted by $RF^{-1}$.

\begin{definition}[Sequential Implementation of an ADT]
 $Z$ is a sequential implementation of  an ADT $A$ w.r.t an abstraction function $AF$, iff for all $op\in Aop$,  $\sigma_z\in ZState, \sigma_a,\sigma_a'\in AState$,   $in \in Input$, $ret\in Output $.
$ AF(\sigma_z)=\sigma_a \,\wedge \, op(\sigma_a,in)=(\sigma_a',ret)
 \Longrightarrow \exists \sigma_z',\; (\sigma_z,in)RF^{-1}(op)(\sigma_z',ret) \wedge  AF(\sigma_z')=\sigma_a'
$.
\end{definition}

A good abstract data type should ensure the important representation independence property.
 We  state  representation independence in terms of observational equivalence---two correct implementations of  an ADT are observationally indistinguishable by  clients of the ADT.
  Application of the definition requires a specific interpretation of what the observable
behaviors really mean.
Client programs access an implementation of an ADT only through the ADT interface. Thus, the  states of implementations of an ADT are unobservable by clients.
In this paper, we take  traces of client programs (i.e. client-side traces) as  observable behaviors.

\begin{definition}[Observational Equivalence]
For an  ADT $A$  and its implementation $Z$ w.r.t the abstraction function $AF$, a client program $P$, the two programs $P(Z)$ and $P(A)$ are observationally equivalent iff  for any initial state $\sigma_{c}$, any well-formed state $\sigma_{z}$,
$ \mathcal{MT}[\kern-0.15em[ P(A)(\sigma_{c},AF(\sigma_{z})) ]\kern-0.15em]=\mathcal{MT}[\kern-0.15em[ P(Z)(\sigma_{c},\sigma_{z}) ]\kern-0.15em]$.
\end{definition}

The following theorem states that when the methods of ADTs are called within their domains, data abstraction implies observational equivalence. The detailed proof is included in the Appendix Section D.

\begin{theorem}
If $Z$ is a sequential implementation of an ADT $A$  then for any client program $P$, if all methods of $A$  are called
 within their domains, then $P(Z)$ and $P(A)$ are observationally equivalent.
\end{theorem}

According to  Definition 12, outside the domain of an abstract method, the corresponding concrete method is  free to do anything, including crashing the program,   returning a correct or incorrect value, or throwing   exceptions. Thus  $P(Z)$ and $P(A)$ can have different behaviors outside  domains of the methods.
Generally, it is the responsibility of clients to ensure that these preconditions hold.

\subsection{Data Abstraction For Concurrent Objects}

A concurrent object  can be viewed as a concurrent implementation of an ADT.
What does it mean for a concurrent object to be an implementation of an ADT?
Like in the sequential setting,  the criterion for relating an ADT  and its  concurrent implementation  should ensure the representation independence property.
Linearizability is not sufficient to capture the property because it only ensures observational refinement, not observational equivalence.
Thus, new observable behaviours can be introduced when clients replace a linearizable object with its corresponding ADT to simplify reasoning about their programs.
For example, a specification of queue can be given as follows:
\[Enqueue(seq,x)=(seq^\smallfrown x, \varepsilon)\]
\[
Dequeue(seq)=
\begin{cases}
(seq',y), &  if \quad seq=y^\smallfrown seq';  \\
(seq, EMPTY),&  if \quad sep= empty;
\end{cases}
\]
Here  $seq$ denotes a sequence, the notation $ \varepsilon$  indicates that a method does not return values. Henzinger et al. [1] show that the HW queue is linearizable with respect to the specification.
Consider the following program:
\[\rm{Enqueue(\textnormal{`}c\textnormal{'})}\;\parallel \; \rm{y=Dequeue()}\]
If the program uses the HW queue, the final value of the variable $y$ is $c$;
if the program uses the abstract queue, the final value of the variable y is $c$ or $empty$.

We present a correctness criterion for a  concurrent implementation of an ADT,  which  is the combination of
 general linearizability and data
abstraction and can ensure observational equivalence.
\begin{definition}[Concurrent Implementation of an ADT]
 A concurrent object $Z$ is a concurrent implementation of an ADT $A$ w.r.t an abstraction function $AF$,  iff \\
1. $Z$ is a sequential implementation of $A$ w.r.t $AF$,\\
2. $Z$ is  linearizable w.r.t. $A$   and
  for every terminating execution  $ (\sigma_z,H_z,{\sigma_z}')$ of $Z$ starting from a well-formed initial state $\sigma_z $, there exists a terminating execution  $(AF(\sigma_z), H_a, AF({\sigma_z}'))$ of $ A$, such that $H_z\sqsubseteq H_a$.\\
\end{definition}

\begin{theorem}
If a concurrent object $Z$ is a concurrent implementation of an ADT $A$ then for  any client program $P$, any well-formed initial state of $Z$, $P(Z)$ and $P(A)$ are observationally  equivalent.
\end{theorem}

The proof for the theorem appears in Appendix Section E. Note that the  observable behaviors  (i.e. the client-side traces)  do not include the traces generated by  divergences of concurrent objects.
In practice, it is the responsibility of  clients to exclude the undesirable
behaviors by ensuring termination of the called methods of concurrent objects  in terms of their progress properties
 and fair assumption.

Data abstraction in the concurrent setting implies atomicity abstraction \textemdash one which enables clients to reason about the operations of concurrent objects as if they occur in a single atomic step. Thus,  for a concurrent object, we refer to such an ADT specification as its abstract atomicity specification.

\subsection{ The Relationship between Atomicity and Abstract Atomicity }

A concurrent object can implement multiple different
ADTs. For example in Appendix Section H,  we show that the MS lock-free queue is not only an implementation of two different queue data types but also an implementation of a multiset data type.
Different abstractions are
suited to different kinds of applications.
It is a challenging problem to prove that a concurrent object  is a concurrent implementation of an ADT,
so clients do not want to have to reverify the implementations each time.

The following theorem shows that for a strict linearizable concurrent object, its sequential specification can serve as ``maximal" atomicity abstraction---for a strict linearizable object $Z$, in order to verify whether $Z$  is a concurrent implementation of an ADT $A$ , it is sufficient to check whether its sequential specification  satisfies the $A's$ specification. The proof for the theorem appears in Appendix Section F.
Thus, for a strict linearizable concurrent object, the challenging problem can reduced to the simpler problem of reasoning about sequential behaviors of the concurrent object.

\begin{theorem}
For a strict linearizable concurrent object $Z$, if for any ADT $A$, $Z$ is a sequential implementation of  $A$ and $\forall op\in Zop,  \sigma_z\in ZState,    in \in Input. (\sigma_z,in)\in dom(op) \Longrightarrow (AF(\sigma_z),in)\in dom(RF(op)) $, then $Z$ is also a concurrent implementation of  $A$.
\end{theorem}

The following theorem can help establish strict linearizability in terms of abstract atomicity. The proof for the theorem appears in Appendix Section G. We show that the MS lock-free queue is strict linearizable in terms of the theorem in Appendix Section H.

\begin{theorem}
 A concurrent object $Z$ is strict linearizable  if there exists an  ADT $A$, such that $Z$ is a  concurrent  implementation  of $A$ w.r.t an injective abstraction function.
 \end{theorem}

\section{Related Work and Conclusion  }
\textbf{Related Work } Strict linearizability is a stronger consistency than general linearizability.
Sequential consistency [29], and quiescent consistency [8,14], as well as relaxed forms of linearizability like quasi linearizability [6] and parameterised linearizability [7],  k-linearizability [12], eventual consistency [13] are weaker consistency conditions than general linearizability, and cannot provide stronger guarantees than strict linearizability.

Several previous works [2,4,26,23] have presented atomicity notions based on serializability (conflict-serializability or view-serializability) and use Lipton's theory of reduction [16,22] as a key technique to prove atomicity.
The correctness criteria are sometimes too restrictive because violations of serializability at the load/store instruction level may not necessarily mean conflicts at the higher, more ``semantic'' level.
 Our notion of strict linearizability, in contrast, defines atomicity for concurrent objects at the sequential specification level.
\\
\textbf{Conclusion } This paper presents a notion of strict linearizability and goes on to show its several important properties which general linearizability cannot capture.
This paper also presents a correctness criterion
 for a  concurrent implementation of an ADT,  which  is the combination of
 general linearizability and data
abstraction and can ensure observational equivalence.
We investigate its relationship with strict linearizability.

\appendix

\section{}
In the Appendix, proofs are
written in a hierarchically structured style as advocated by
Lamport [27].

\textbf{Proposition 2.}\;\;
$H_1 \sqsubseteq H_2 \wedge  H_2 \sqsubseteq H_3 \Longrightarrow H_1 \sqsubseteq H_3$
\begin{proof}
1 $\forall t. H_1\lceil t = H_3\lceil t$;\\
proof: $\forall t. H_1\lceil t = H_2\lceil t$; and $\forall t. H_2\lceil t = H_3\lceil t$;\\
2 Assume: \\
 \quad  (a)  $  \nu_1 : \{1, \ldots, |H_1|\} \to \{1, \ldots, |H_2|\}$ such
that $  \forall i.  H_1(i)= H_2(\nu_1 (i))$ and
$\forall i,j. i < j \wedge H_1(i) \in resAct \wedge H_1(j)\in invAct \Longrightarrow \nu_1(i) < \nu_1(j)$.\\
 \quad  (b)  $  \nu_2 : \{1, \ldots, |H_2|\} \to \{1, \ldots, |H_3|\}$ such
that $  \forall i.  H_2(i)= H_3(\nu_2 (i))$ and
$\forall i,j. i < j \wedge H_2(i) \in resAct \wedge H_2(j)\in invAct \Longrightarrow \nu_2(i) < \nu_2(j)$.\\
proof: by the definition of linearizability relation.\\
3  Let $  \nu_3 $ be a bijection $  \{1, \ldots, |H_1|\} \to \{1, \ldots, |H_3|\}$ such that $  \forall i. H_1(i)=H_3(\nu_2(\nu_1(i)) $. Then
$\forall i,j. i < j \wedge H_1(i) \in resAct \wedge H_1(j)\in invAct \Longrightarrow \nu_2(\nu_1(i)) < \nu_2(\nu_1(j))$.
\\
proof:$ \forall i,j. i < j \wedge H_1(i) \in resAct \wedge H_1(j)\in invAct \Longrightarrow H_2(\nu_1 (i))\in resAct \wedge  H_2(\nu_1(j))\in invAct \wedge \nu_1(i) < \nu_1(j)$.
Thus  $\forall i,j. i < j \wedge H_1(i) \in resAct \wedge H_1(j)\in invAct \Longrightarrow \nu_2(\nu_1(i)) < \nu_2(\nu_1(j))$.\\
4 Q.E.D.\\
proof: By 1 and 3.
\end{proof}

\section{}
Because the states of a client program are disjoint from the ones of a concurrent object,
we can divide  an execution $\pi=(\sigma_{c0},(\sigma_{z0},\phi))$ $\xrightarrow{e_1}
 (\sigma_{c1},(\sigma_{z1},u_1)),
\cdots,$ $\xrightarrow{e_n}(\sigma_{cn},(\sigma_{zn},u_n))$ of a program $ P(Z)$
into an execution $\pi_c$ of the client program  $P$ and an execution $\pi_z$ of the object $Z$
 as follows:
\\
\[\pi_c =(\sigma_{c0})\xrightarrow{b_1},\cdots,\xrightarrow{b_n}( \sigma_{cn}),
\]
\[\pi_z =(\sigma_{z0},\phi)\xrightarrow{c_1},\cdots,\xrightarrow{c_n}(\sigma_{zn},u_n), \]
where $ tr(\pi_c )=tr(\pi)\lceil A_c$ and $tr(\pi_z)=tr(\pi)\lceil A_z$.

To simplify our notation, we use the abbreviation\\
 $(\sigma_{z})$ $\xrightarrow{\langle inv(op, n),ret(v) \rangle}$ $(\sigma_z')$
to describe the atomic  execution of the operation $op$ staring from the initial state $\sigma_{z}$ with an input $n$ and terminating in the state $\sigma_z'$ with an output $v$.

For two histories $ H$ and $ H'$, if $ H \sqsubseteq H' $, we can establish a bijective function $\mathbb{F}$ mapping between operations in $H$ and $H'$;  an operation $o$ in $H$  is mapped to  an operation $o'$ in $H'$ by $\mathbb{F}$ if for all thread $t$, $ H\lceil t (i)=inv  \Longleftrightarrow H'\lceil t (i)=inv' \wedge F(OP(inv))=OP(inv')$.
\\
\textbf{Theorem  6.}\;\;
A concurrent object $Z$ is strict linearizable
iff for any client program $P$, any initial state $\sigma_{c}$ of $P$, any well-formed state $\sigma_{z}$ of  $Z$,\\
1. $ \mathcal{MT}[\kern-0.15em[ P(Z)(\sigma_{c},\sigma_{z}) ]\kern-0.15em]=\mathcal{MT}[\kern-0.15em[ P(Ato\_Z)(\sigma_{c},\sigma_{z}) ]\kern-0.15em] $\\
2.
 $ \mathcal{MS}[\kern-0.15em[ P(Z)(\sigma_{c},\sigma_{z}) ]\kern-0.15em]=\mathcal{MS}[\kern-0.15em[ P(Ato\_Z)(\sigma_{c},\sigma_{z}) ]\kern-0.15em]$
\begin{proof}
By Lemma 1 and Lemma 4.
\end{proof}

\begin{lemma}
A concurrent object $Z$ is strict linearizable
then for any client program $P$, any initial state $\sigma_{c}$ of $P$, any well-formed state $\sigma_{z}$ of  $Z$,\\
1. $ \mathcal{MT}[\kern-0.15em[ P(Z)(\sigma_{c},\sigma_{z}) ]\kern-0.15em]=\mathcal{MT}[\kern-0.15em[ P(Ato\_Z)(\sigma_{c},\sigma_{z}) ]\kern-0.15em] $\\
2.
 $ \mathcal{MS}[\kern-0.15em[ P(Z)(\sigma_{c},\sigma_{z}) ]\kern-0.15em]=\mathcal{MS}[\kern-0.15em[ P(Ato\_Z)(\sigma_{c},\sigma_{z}) ]\kern-0.15em]$
\end{lemma}
\begin{proof}
By Lemma 2 and Lemma 3.
\end{proof}

\begin{lemma}
For any client program $P$, any initial state $\sigma_{c}$ of the client program, any well-formed state $\sigma_{z}$ of the object $Z$ :\\
\indent (1)$ \mathcal{MT}[\kern-0.15em[ P(Z)(\sigma_{c},\sigma_{z}) ]\kern-0.15em]\subseteq\mathcal{MT}[\kern-0.15em[ P(Ato\_Z)(\sigma_{c},\sigma_{z}) ]\kern-0.15em] $
\\
\indent  (2)
 $ \mathcal{MS}[\kern-0.15em[ P(Z)(\sigma_{c},\sigma_{z}) ]\kern-0.15em]\subseteq\mathcal{MS}[\kern-0.15em[ P(Ato\_Z)(\sigma_{c},\sigma_{z}) ]\kern-0.15em]$
 \end{lemma}

\begin{proof}\\
 1  For any terminating execution $\pi$ of $P(Z)$,
 there exists an execution $\pi'$ of $P(Ato\_Z)$ such that:\\
   (1) $\pi'$  and $\pi$  have the same client-side traces.\\
   (2) $\pi'$  and $\pi$  have the same final states.\\
 \indent  proof:
   1.1 Assume: For any terminating execution $\pi$ of $P(Z)$:\\
 $\pi =(\sigma_{c0},\sigma_{z0})
 \xrightarrow{\lambda}$$(\sigma_{cn},\sigma_{zn})$. By separating the execution $\pi$, we can get an execution
$\pi_c=(\sigma_{c0})\xrightarrow{\gamma}( \sigma_{cn})$ of  the client program,
where \,$\gamma=\lambda\lceil{A_c}$.\\
To prove that the lemma holds, we need to prove that there exists an execution
${\pi}'$$ =$$(\sigma_{c0},\sigma_{z0})$ $\xrightarrow{\beta}(\sigma_{cn},\sigma_{zn})$ of $P(Ato\_Z)$,
such that $\lambda\lceil{A_c}$ $=$ $\beta\lceil{A_c}$.
\\
 \indent 1.2  There exists an execution
$\pi_a$ $ =$ $\sigma_{z0}$ $\xrightarrow{\langle inv(op_1', n_1),ret(v_1) \rangle}$ $\sigma_{z1}', \cdots,\\
\xrightarrow{\langle inv(op_n', n_n),ret(v_n) \rangle}$$\sigma_{zn}$ of $Ato\_Z$, such that $H(\pi )$ $\sqsubseteq$ $H(\pi_a )$.
\\ \indent proof: By the  definition of strict linearizability.
 Let the function $\mathbb{F}$ map every operation $op_i$ in $H(\pi)$  to every operation $op_i'$ in $H(\pi_a)$
\\
 \indent 1.3  For every action  $\langle inv(op_i',n_i), ret(v_i)\rangle $ in $\pi_a$,
there exists two atomic actions  in $\pi_c $: one of which  is argument evaluation  of the operation $op_i$ in $\pi$ (denoted by $e_i$), the other of which  is assignment of the return value of the operation $op_i$ to a client's variable (denoted by $x_i:=ret_i$) such that the value of  $e_i$ is $n_i$ and $v_i=ret_i$.
\\ \indent  proof: Since $H(\pi )\sqsubseteq_{Hlin}H(\pi_a )$, the two operations $op_i'$ and $op_i$  have the same argument values and return values.\\
 \indent 1.4  Every atomic action $\langle inv(op_i', n_i), ret(v_i)\rangle $ in $\pi_a $,
can be inserted between $e_i$ and $x_i:=ret_i$ in $\pi_c$ and preserves the real time order in $\pi_a$.
\\ \indent
proof: By induction on n.\\
\indent  1.4.1 Base case: when n=1, 1.4 is true.\\
 \indent \indent proof: trivial.\\
\indent 1.4.2 Inductive hypothesis: when n=k, 1.4 is true.\\
\indent 1.4.3 Inductive step: when n=k+1, 1.4 is true.\\
\indent\indent 1.4.3.1 By inductive hypothesis, to prove 1.4.3, we need to prove that \\  $\langle inv(op_{(k+1)}', n_{(k+1)}), ret(v_{(k+1)})\rangle$ can be inserted between $e_{(k+1)}$ and $x_{(k+1)}:=ret_{(k+1)}$ and behind $\langle inv(op_k', n_k), ret(v_k)\rangle $. \\
 \indent\indent  1.4.3.2 $e_k<_a x_{(k+1)}:=ret_{(k+1)}$ in  $\pi_c$.\\
\indent\indent proof:
Assume $x_{(k+1)}:=ret_{(k+1)}<_a e_k$, we can get (a) $op_{(k+1)}<_o op_k$ in $\pi$.
Since $op_k' <_o op_{(k+1)}'$ in $\pi_a $ and  $H(\pi )\sqsubseteq H(\pi_a )$,  we can get (b) $op_{(k+1)}\nless_o op_k$. Thus, a contradicts b,  the original assumption must have been wrong.\\
\indent\indent  1.4.3.3 All possible happened-before total orders on  $e_k$, $x_k:=ret_k $,  $e_{(k+1)}$, $x_{(k+1)}:=ret_{(k+1)}$ are shown as follows:\\
(1)\,$e_{(k+1)}<_a e_k<_a x_{(k+1)}:=ret_{(k+1)}<_a x_k:=ret_k $ \\
(2)\,$e_k<_a e_{(k+1)}<_a x_{k+1}:=ret_{(k+1)}<_a x_k:=ret_k $ \\
(3)\,$e_{(k+1)}<_a e_k<_a x_k:=ret_k<_a x_{(k+1)}:=ret_{(k+1)}$ \\
(4)\,$e_k<_a e_{(k+1)}<_a x_k:=ret_k<_a x_{(k+1)}:=ret_{(k+1)}$ \\
(5)\,$e_k<_a x_k:=ret_k<_a e_{(k+1)}<_a x_{(k+1)}:=ret_{(k+1)}$.\\
\indent \indent \indent proof: By 1.4.3.2 and  $e_{(k+1)}<_a x_{(k+1)}:=ret_{(k+1)}$.
\\
\indent\indent  1.4.3.4 Q.E.D.
\\
\indent \indent \indent proof: For any happened-before order in 1.4.3.3, we  can  insert  $\langle inv(op_k', n_k), ret(v_k)\rangle$  between $e_k$ and  $x_k:=ret_k$,  insert  $\langle inv(op_{(k+1)}',$
$ n_{(k+1)}), ret(v_{(k+1)})\rangle$  between $e_{(k+1)} $ and  $x_{(k+1)} :=ret_{(k+1)} $, and  preserve the happened-before order $op_k' <_o op_{(k+1)}'$.\\
\indent 1.4.4.  Q.E.D.
\\
\indent \indent Proof: By 1.4.1 and 1.4.2 and 1.4.3.\\
 \indent 1.5 Let  $\xi$ be the trace by inserting every $\langle inv(op_i', n_i), ret(v_i)\rangle $ in $\pi_a $ for  $ i=1\ldots n $ into $tr(\pi_c )$ as 1.4. The execution
 $\pi_\tau$$ =$$(\sigma_{c0},\sigma_{z0})$ $\xrightarrow{\xi}$$(\sigma_{cn},\sigma_{zn})$\ is feasible.\\
\indent proof: By separating the execution $\pi_\tau$, we can get  two feasible executions $\pi_c$ and $\pi_a$.
By our semantics a state of a client program is disjoint from that
of a concurrent object, thus  $\pi_\tau$ is a feasible execution.\\
 \indent 1.6 $\pi_\tau$ is an execution of $P(Ato\_Z)$.\\
\indent proof: By the constructing process of $\pi_\tau$,
$\pi_\tau$ is an execution of the program which is  the same  as $P(Z)$ except the statement $x_i:=Z.op_i(e_i)$, which is replaced by $x_i:=Ato\_Z.op_i'(e_i)$.\\
 \indent 1.7. Q.E.D.\\
\indent proof: By 1.1, 1.5, 1.6.
\\
2 For any divergent execution $\beta$ of $P(Z)$
caused by  divergence of the client program,
 there exists an execution $\beta'$ of $P(Ato\_Z)$ such that:
 $\beta'$  and $\beta$  have the same client-side traces.\\
   proof:
   Note that the definition of the client-side traces only considers divergence caused
by a client program. Thus, for a divergent execution of $P(Z)$, we need only
consider the case: a divergent execution of $P(Z)$ where the execution of $Z$
is finite. The proof for the case is similar to the above one.\\
3 Q.E.D.\\
 \indent  proof:  By 1 and 2.
\end{proof}

\begin{lemma}
 For any client program $P$, any initial state $\sigma_{c}$ of the client program, any well-formed state $\sigma_{z}$ of the object $Z$:\\
 \indent (1)$ \mathcal{MT}[\kern-0.15em[ P(Ato\_Z)(\sigma_{c},\sigma_{z}) ]\kern-0.15em]\subseteq\mathcal{MT}[\kern-0.15em[ P(Z)(\sigma_{c},\sigma_{z}) ]\kern-0.15em] $\\
 \indent (2)
 $ \mathcal{MS}[\kern-0.15em[ P(Ato\_Z)(\sigma_{c},\sigma_{z}) ]\kern-0.15em]\subseteq\mathcal{MS}[\kern-0.15em[ P(Z)(\sigma_{c},\sigma_{z}) ]\kern-0.15em]$
 \end{lemma}
 \begin{proof}
  \indent Trivial.
  \end{proof}

\begin{lemma}
For any client program $P$, any initial state $\sigma_{c}$ of $P$, any well-formed state $\sigma_{z}$ of  $Z$, if \\
1. $ \mathcal{MT}[\kern-0.15em[ P(Z)(\sigma_{c},\sigma_{z}) ]\kern-0.15em]=\mathcal{MT}[\kern-0.15em[ P(Ato\_Z)(\sigma_{c},\sigma_{z}) ]\kern-0.15em] $\\
2.
 $ \mathcal{MS}[\kern-0.15em[ P(Z)(\sigma_{c},\sigma_{z}) ]\kern-0.15em]=\mathcal{MS}[\kern-0.15em[ P(Ato\_Z)(\sigma_{c},\sigma_{z}) ]\kern-0.15em]$\\
 then $Z$ is strict linearizable.
\end{lemma}

\begin{proof}\\
1.  For any terminating execution $\pi:(\sigma_z,H,\sigma_z')$ of the object $Z$,  there exists a sequential execution
$(\sigma_z,H',\sigma_z')$  of the object $Z$ such that
$H\sqsubseteq H'$.\\
\indent 1.1 To simplify writing,
we assume that a thread  invokes a method of $ Z$ at most once in $\pi$.
We now construct a program $P(Z)$:
$ s_1; x_1=Z.m_1(e_1);s_1';\!\parallel \!\cdots  s_i; x_i=Z.m_i(e_i);s_i'\!\parallel\!\cdots  s_n; x_n=Z.m_n(e_n);s_n'  $
where for each $i$, $m_i(e_i)$ is a method called by the thread $i$ in $\pi$ and $e_i$ is an argument of the method $m_i$; $s_i$ and $s_i'$ are atomic regions and and $x_i$ is a variable of the client program $P$. Let $\sigma_c $ be the client initial state of $P(Z)$ where all variables of the client program are null.
\\
\indent  1.2 There exists an execution $\pi'$  of $P(Z)$ starting from the initial state $(\sigma_c,\sigma_z)$ such that: \\
 $tr(\pi)=tr(\pi')\lceil A_z $ ;
 the action  $x_i=ret_i$ and $s_i'$is executed  immediately after the method $m_i$ returns where $ret_i$ is  the return value of $m_i$;
 the method $m_i$ is  immediately invoked after $s_i$ finishes. Let $(\sigma_c',\sigma_z'')$ be the final states of $\pi'$.\\
 \indent proof: The trace of  $\pi'$  can be obtained by inserting $x_i=ret_i$ and $s_i'$ after the returning action of $m_i$ and inserting $s_i$ before the invocation action of $m_i$ in the trace of $\pi$. Obviously, the execution $\pi'$  is feasible.\\
\indent  1.3 $m_{(i+1)}>m_i\Longrightarrow s_{(i+1)}>x_i=ret_i $ in $\pi'$ for $i=1,\ldots,n $.\\
  \indent proof: by the constructing process of $\pi'$.\\
\indent  1.4  Consider the program $P(Ato\_Z)$:
 $ s_1; x_1=Ato\_Z.\langle m_1(e_1)\rangle;s_1' \!\parallel \!\cdots  s_i; x_i=Ato\_Z.\langle m_i(e_i) \rangle \!\parallel\!\cdots  s_n; x_n=Ato\_Z.\langle m_n(e_n)\rangle $ where $\langle m_i(e_i)\rangle $ is an atomic version of the method $m_i(e_i)$.
For any terminating execution $\pi'' $
   of the program $P(Ato\_Z)$  starting from  $(\sigma_c,\sigma_z)$, if
 $tr(\pi')\lceil A_c=tr(\pi'')\lceil A_c $
 then $H(tr(\pi')\lceil A_z)\sqsubseteq H(tr(\pi'')\lceil A_z)$.\\
 \indent \indent 1.4.1 $\forall t.  H(\pi')\lceil t =H(\pi'')\lceil t $\\
 proof:
 Since $tr(\pi')\lceil A_c=tr(\pi'')\lceil A_c $,
 the return value $  of  \langle m_i(e_i) \rangle  $ in $ \pi''$ is equal to that of $ m_i(e_i)$ in $\pi'$ for $i=1,\ldots,n $.\\
 \indent \indent 1.4.2   For any two operations
  $m_{y1}$, $m_{y2} $ in $H(tr(\pi')\lceil A_z)$, and $\langle m_{y1}\rangle$, $\langle m_{y2} \rangle$ in $H(tr(\pi'')\lceil A_z)$, $m_{y2}>m_{y1} \Longrightarrow \langle m_{y2}\rangle > \langle m_{y1}\rangle$\\
 proof: by 1.3, $m_{y2}>m_{y1}  \Longrightarrow s_{y2}>x_{y1}=ret_{y1} $.  In $\pi''$, $ \langle m_{y1}\rangle$ finishes  before $ x_{y1}=ret_{y1}$ and $ \langle m_{y2}\rangle$ starts after $ s_{y2}$. Thus
 $ \langle m_{y2}\rangle > \langle m_{y1}\rangle$.
 \\
 \indent \indent 1.4.3 Q.E.D.\\
  proof: By 1.4.1 and 1.4.2.\\
\indent  1.5  $tr(\pi')\lceil A_c$ is only a client trace can lead to the client state $ \sigma_c'$.
 \\
 proof:
 For each atomic region $s_i$ or $s_i'$, we can construct it
 by the following rule:
 There is a variable $y_i$ (or $y_i'$ ) at the atomic region $s_i$ (or $s_i'$ ); the variables $y_i$ (or $y_i'$ ) is assigned to different values when $s_i$ (or $s_i'$ ) is executed in different orders.
A key method for doing this is that the atomic region $s_i$ can determine whether the other actions have been executed in terms of values of their corresponding variables. For example, the code of the atomic region $s_1$ can be defined as follows:\\
$\langle if (x_2=null \;and \;\dots\; and\; x_n=null\\
and\; y_2=null\; and \;\dots \;and\; y_n=null \\
and\; y_2'=null\; and \;\dots \;and\; y_n'=null)\\
y_1=case1\\
else \;if (\dots)\\
y_1=case2\\
\dots \rangle$\\
The trace $tr(\pi')\lceil A_c$ has  the following form:\\
$\dots s_i\dots x_i=ret_i,s_i'\dots s_j\dots x_j=ret_j,s_j'\dots $\\
Changing the order of the trace affects at least a position of an atomic region.
\\
\indent 1.6 There exists an execution $\pi''' $ of the program $P(Ato\_Z)$  starting from  $(\sigma_c,\sigma_z)$
 such that the final state is $(\sigma_c',\sigma_z')$.
 \\
 proof: By the second condition of the lemma.\\
 \indent  1.7 Q.E.D.\\
 By 1.5 and 1.6, $tr(\pi')\lceil A_c=tr(\pi''')\lceil A_c$.
 Thus, by 1.4, $H(\pi')\sqsubseteq H(\pi''')$.
\\
2  For any  non-terminating execution $\varphi$ of the object $Z$,  there exists a sequential execution
$\varphi'$  of the object $Z$ and a history $h_c\in Compl(H(\varphi))$ such that
$h_c\sqsubseteq H(\varphi') $.
\\
\indent 2.1
To simplify writing,
we assume that a thread  invokes a method of $ Z$ at most once in $\varphi$.
 We now construct a program $P(Z)$:
$ s_1; x_1=Z.m_1(e_1)\!\parallel \!\cdots  s_i; x_i=Z.m_i(e_i)\!\parallel\!\cdots  s_n; x_n=Z.m_n(e_n) \!\parallel\!s_{abort} $
where $m_i(e_i)$ is a method called by the thread $i$ in $\pi$ and $e_i$ is an argument of the method $m_i$; $s_i$  is an atomic statement and and $x_i$ is a variable of the client program $P$. $s_{abort}$ is  an atomic statement and its execution will lead to the error state $abort$.
 There exists an execution $\xi$ of $P(Z)$ such that: \\
 (1) $tr(\varphi)=tr(\xi)\lceil A_z $.\\
(2)  The method $m_i$ is  immediately invoked after $s_i$ finishes;
The action  $x_i=ret_i$ is executed  immediately after the method $m_i$ returns where $ret_i$ is  the return value of $m_i$ if for each $i$, $m_i$ returns;
\\
 \indent proof: The trace of  $\xi$  can be obtained by inserting $x_i=ret_i$ into  $\pi$ at the position after the returning action of $m_i$ and inserting $s_i$ at the position before the invocation action of $m_i$. Obviously, the execution $\xi$  is feasible.\\
\indent 2.2  For any two complete methods $ m_{(i+1)}, m_i$ in $\pi'$, $m_{(i+1)}>m_i\Longrightarrow s_{(i+1)}>x_i=ret_i $ for $i=1,\ldots,n $.\\
  \indent proof: by the constructing process of $\xi$.\\
\indent 2.3  Consider the program $P(Ato\_Z)$ :
 $ s_1; x_1=Ato\_Z.\langle m_1(e_1)\rangle \!\parallel \!\cdots  s_i; x_i=Ato\_Z.\langle m_i(e_i) \rangle \!\parallel\!\cdots  s_n; x_n=Ato\_Z.\langle m_n(e_n)\rangle $ where $\langle m_i(e_i)\rangle $ is an atomic version of the method $m_i(e_i)$.
 There exists an execution $\xi' $ of the program $P(Ato\_Z)$ such that
 $tr(\xi)\lceil A_c=tr(\xi')\lceil A_c $ \\
 proof:
 By  client-side traces equivalence between $P(Z)$ and $P(Ato\_Z)$.\\
\indent 2.4  For any  complete method $m_i(e_i)$ in $\xi$,
the return value of the method is equal to that of $\langle m_i(e_i)\rangle$ in $\xi'$.\\
 proof:
By $tr(\xi)\lceil A_c=tr(\xi')\lceil A_c $.\\
\indent 2.5   For any two complete methods $m_{y1}$, $m_{y2} $ in $\pi'$,  $m_{y2}>m_{y1} \Longrightarrow \langle m_{y2}\rangle > \langle m_{y1}\rangle$\\
 proof: by 2.2, $m_{y2}>m_{y1}  \Longrightarrow s_{y2}>x_{y1}=ret_{y1} $.  In $\xi'$, $ \langle m_{y1}\rangle$  must finish  before $ x_{y1}=ret_{y1}$ and $ \langle m_{y2}\rangle$ must start after $ s_{y2}$. Thus
 $ \langle m_{y2}\rangle > \langle m_{y1}\rangle$.
 \\
\indent 2.6  Q.E.D.\\
 We construct a completion  $h_c$ of $H(\varphi)$ by the following step:\\
 For a pending invocation event of a method in $H(\xi)$,
  if
 the method is not in $ H(\xi')$,
 then we delete the pending invocation event; Otherwise, we add the same response events of the method as that of the method in $H(\xi')$.
 By 2.4 and 2.5 and the constructing process of $ h_c$,
 $h_c\sqsubseteq H(\xi') $
 \\
3  Q.E.D.\\
proof: By 1 and 2.
\end{proof}

\textbf{Theorem  7.}\;\;
For a strict linearizable object $Z$ with a well-formed initial state,  any client program $P$,
$P(Z)$ and $P(Ato\_Z)$ have the same client-side traces and final states even in the relaxed program model above.
\begin{proof}
For any program $P(Z)$,
we encapsulate the shared address read/write actions of $P$ into methods,
and obtain the object $Z'$ by adding the new methods into $Z$,
the client program $P'$ by replacing the the shared address read/write actions
of $P$ with the new methods.
The concurrent history generated by $P'(Z')$ is of the form
\[CH= (CH_1)^\frown (SH_1)^\frown (CH_2)\frown(SH_2)\dots\]
\[or\;\;  CH= (SH_1)^\frown (CH_1)^\frown(SH_2)^\frown (CH_2)\dots\]
where
for each $i$, $SH_i$  denotes the sequential history is generated by the new methods which
encapsulate the compatible actions satisfying the first restriction;
$CH_i$ denotes  the concurrent history is generated by the old methods   and
 the new methods which
encapsulate the compatible actions satisfying the second restriction;
$^\frown$ denotes the concatenation of two histories;

For each $CH_i$, there exists a linearization $CH_i'$ of $CH_i$, such that
the concurrent execution which generate $CH_i$ and the sequential execution which generate
$CH_i'$ have the same final state of $Z$.
Thus,
$SH=CH_1'^\frown SH_1\dots $ or $SH=SH_1^\frown CH_1' \dots $ is a  linearization of $CH$ and the concurrent execution
which generate $CH$ and the sequential execution which generate
$SH$ have the same final state of $Z$.
Similar to the proof of Theorem 6, $P'(Z')$ and $P'(Ato\_Z')$ have the same final states and the same traces generated by
the clint program $P'$ and the new methods.
Each execution of $P'(Z')$ can correspond to an execution of $P(Z)$, and vice versa.
Each execution of $P(Ato\_Z)$ can correspond to an execution of $P'(Ato\_Z')$, and vice versa.
Thus, $P(Z)$ and $P(Ato\_Z)$ have the same final states and the same client-side traces.

\end{proof}

\section{}

\textbf{Theorem  10.}\;\;
For a strict linearizable and  purely-blocking  object $Z$, any client program $P$, $P(Z)$ diverges  iff  $P(Ato\_Z)$ diverges.
\begin{proof}
By the following lemma (Lemma 5) and Corollary 8.
\end{proof}

\begin{lemma}
For a strict linearizable and  purely-blocking object $Z$, any client program $P$,
$P(Z)$ diverges by  divergences of $Z$ iff  $P(Ato\_Z)$ diverges by  divergences of  $Ato\_Z$.
\end{lemma}

\begin{proof}
($\Rightarrow$)
Assume: $P(Z)$ does not diverge by e divergences of $Z$.
Let $S$ be a divergent execution of $Ato\_Z$, which is obtained by separating a divergent execution of $P(Ato\_Z)$ by the divergence of  $Ato\_Z$.
Consider such an execution $S'$ of $Z$. $S'$ executes the methods which is finished in $S$ sequentially by the same order as $S$, then call other methods. In terms of the assumption above, all called methods in $S'$ will finish. Since $S$ is divergent, there does not exist a linearization of $S'$, contradicting the fact that $Z$ is strict linearizable.
\end{proof}

\begin{proof}
($\Leftarrow$)
Assume: $P(Ato\_Z)$ does not diverge by  divergences  of $Ato\_Z$.
Let $S$ be a divergent execution of $Z$, which is obtained by separating a divergent execution of $P(Z)$ by the divergence of  $Z$.
In terms of the definition of strict linearizability and purely-blocking,
 there exists a sequential execution
$S'$ of $Z$ and a history $h_c\in Compl(H(S))$ such that
 such that $h_c\sqsubseteq H(S')$, and the final state of $Z$ in $S'$ is the same as that of $Z$ in $S$ (Because pending methods do not change the global states, we call the state of $Z$ the final state of $Z$ in $S$ after all called methods of $S'$ which can finish finish). Let $\sigma_z$ denote the final state of $Z$.
Consider such an execution $S''$ of $Ato\_Z$.  (1) firstly, $S''$ executes the methods of $S'$ sequentially by the same order as $S'$; (2)then, calls other methods. The state of $Z$ is $\sigma_z$ after 1. In terms of the assumption above, there at least exists a method can finish in $\sigma_z$. This contradicts the fact that no pending methods of $S$ in $\sigma_z$ can finish.
\end{proof}

Similar to the proof of Theorem 7, we can show that the theorem still holds for the relaxed program model in Subsection 5.1.

\section{}
\textbf{Theorem 14.}\;\;
If $Z$ is a sequential implementation of an ADT $A$  then for any client program $P$, if all methods of $A$ are called
 within their domains, then $P(Z)$ and $P(A)$ are observationally equivalent.
\begin{proof}
1 For any terminating execution  $\mu$: $ (\sigma_a,H_a,\sigma_a')$ of $A$, there exists a terminating sequential execution  $\mu'$:
 $ (\sigma_z, H_{z},\sigma_z')$ of $Z$, such that $AF(\sigma_z)=\sigma_a $, $AF(\sigma_z')=\sigma_a'$ and  $H_a=H_z$.
 \\
proof: This is proved in the 2 of the proof of Lemma 6 in Appendix Section G.
\\
2
For any terminating execution  $\mu$:
 $ (\sigma_z, H_{z},\sigma_z')$ of $Z$ which is generated in
 $P(Z)$, there exists a  sequential and terminating execution  $\mu'$: $ (\sigma_a,H_a,\sigma_a')$ of $A$, such that $AF(\sigma_z)=\sigma_a $, $AF(\sigma_z')=\sigma_a'$ and  $H_a=H_z$.
\\
proof: By the lemma's hypothesis,  the methods of of $A$ in $P(A)$ are called within their domains. Thus, By 1, we can get 2.
\\
3 Q.E.D.
\\
proof: By 1 and 2. The proof is similar to the one for Lemma 2.
\end{proof}

\section{}

\textbf{Theorem 16.}\;\;
If a concurrent object $Z$ is a concurrent implementation of an ADT $A$ then for  any client program $P$, any well-formed initial state of $Z$,$P(Z)$ and $P(A)$ are observationally  equivalent.
\begin{proof}
1   For any terminating execution  $\mu_a$: $ (\sigma_a,H_a,\sigma_a')$ of $A$, there exists a sequential and terminating execution  $\mu_z$:
 $ (\sigma_z, H_z,\sigma_z')$  of $Z$, such that $AF(\sigma_z)=\sigma_a $, $AF(\sigma_z')=\sigma_a'$ and  $H_a=H_z$.
 \\
 proof: This is proved in the 2 of the proof of Lemma 6 in Appendix Section G.
 \\
 2
For any concurrent execution $\pi_z$ of $Z$ starting from any well-formed state $\sigma_z$,
there exists a legal sequential execution $\pi_a$ of $A$ starting from the state $AF(\sigma_z)$  and a history $h_c \in Compl(H(\pi_z))$  such that $ h_c\sqsubseteq H(\pi_a) $.
\\
proof: By  the definition of concurrent implementation of an ADT (Definition 15).
\\
3 Q.E.D.\\
proof: By 1,2. The proof is similar to the one for Lemma 2.
\end{proof}

\section{}
\textbf{Theorem 17.}\;\;
For a strict linearizable object $Z$, if for any ADT $A$, $Z$ is a sequential implementation of  $A$ and $\forall op\in Zop,  \sigma_z\in ZState,    in \in Input.  (\sigma_z,in)\in dom(op) \Longrightarrow (AF(\sigma_z),in)\in dom(RF(op)) $, then $Z$ is a concurrent implementation of  $A$.
\begin{proof}
1  For any terminating execution  $ \pi_z: (\sigma_z,H,\sigma_z')$ of $Z$, there exists an  execution  $ \pi_a: (\sigma_{a},H_a,\sigma_a')$ of $A$,  such that $ H\sqsubseteq H_a$ and $AF(\sigma_{z})=\sigma_{a} $, $AF(\sigma_z')=\sigma_a'$.

1.1  For any  sequential and terminating execution $\mu:$ $ (\sigma_{z0},H_z,\sigma_{zn})$ of $Z$ , there exists a sequential and terminating execution $\mu':$ $(\sigma_{a0}, H_{a},\sigma_{an})$ of $A$, such that $AF(\sigma_{z0})=\sigma_{a0} $, $AF(\sigma_{zn})=\sigma_{an}$ and  $H_a=H_z$.
 \\
\indent  1.1.1 Assume:
 $\mu=\sigma_{z0}\xrightarrow{inv(op_1,
n_1),ret(v_1)}
(\sigma_{z1})\xrightarrow{inv(op_2,
n_2),ret(v_2)}
\sigma_{z2},\cdots, $  $\xrightarrow{inv(op_n,
n_n),ret(v_n)}$$\sigma_{zn}$
 \\
There exists an execution of $A$,
$\mu'=\sigma_{a0}\xrightarrow{\langle inv(RF(op_1),n_1),ret(v_1) \rangle}
\sigma_{a1}\\
\xrightarrow{\langle inv(RF(op_2), n_2),ret(v_2) \rangle}
\sigma_{a2},\cdots, $  $\xrightarrow{\langle inv(RF(op_n), n_n),ret(v_n) \rangle}$ $\sigma_{an}$ such that
$AF(\sigma_{zi})= \sigma_{ai}$, for $i=0\cdots n$.
\\
proof:
Let $\sigma_{a0}$ be the state such that $AF(\sigma_{z0})= \sigma_{a0}$.
By  the lemma's hypothesis, $(\sigma_{a0},n_1)\in dom(RF(op))$. Thus there exists $\sigma_{a1}$, $ret(v_1)'$
such that  $(\sigma_{a0})\xrightarrow{inv(RF(op_1),
n_1),ret(v_1)'}(\sigma_{a1})$.
 By  Definition 15, $ret(v_1)'=ret(v_1)$ and $AF(\sigma_{z1})= \sigma_{a1}$.
 Thus $\sigma_{a0}\xrightarrow{inv(RF(op_1),
n_1),ret(v_1)}\sigma_{a1}$.
By similar reasoning,
 there exists  $\sigma_{a(i+1)}$ such that
$\sigma_{ai}\xrightarrow{inv(RF(op_{(i+1)}),
n_{(i+1)}),ret(v_{(i+1)})}\sigma_{a(i+1)}$ and $AF(\sigma_{z(i+1)})= \sigma_{a(i+1)}$,
for each $i=1\cdots n-1$.
\\
1.1.2 Q.E.D.\\
\indent \indent proof: By 1.1.1.
\\
 1.2
For the terminating execution  $ \pi_z: (\sigma_z,H,\sigma_z')$ of $Z$, there exists a sequential execution  $ \pi_z': (\sigma_z, H',\sigma_z')$  of $Z$   such that $ H\sqsubseteq H'$.\\
  \indent proof: Since $Z$ is strict linearizable.
  \\
  1. 3. Q.E.D.\\
 \indent proof:
 By 1.1,
 for the sequential execution  $\pi_z': (\sigma_{z}, H',\sigma_z')$ of $Z$,
there exists an execution
 $ \pi_a: (\sigma_{a}, H_a,\sigma_a')$  of $A$, such that $AF(\sigma_{z})=\sigma_{a} $, $AF(\sigma_z')=\sigma_a'$ and  $H_a=H'$.
 Since  $H_a=H'$ and $ H\sqsubseteq H'$, $ H\sqsubseteq H_a$.
 Thus,
for the concurrent execution of $Z$, $ \pi_z: (\sigma_z,H,\sigma_z')$, there exists an  execution of $A$, $ \pi_a: (\sigma_{a},H_a,\sigma_a')$   such that $ H\sqsubseteq H_a$ and $AF(\sigma_{z})=\sigma_{a} $, $AF(\sigma_z')=\sigma_a'$.
\\
2  $Z$ is linearizable w.r.t. $A$.
\\
2.1  For any terminating execution $\varphi$ of $Z$ starting from any well-formed state $\sigma_z$,
there exists a  sequential and terminating execution $\varphi'$ of $Z$, and a history $h_c \in Compl(H(\varphi))$  such that $ h_c\sqsubseteq H(\varphi') $.
\\
proof: Since $Z$ is strict linearizable.
\\
2.2  There exists a sequential and terminating execution $\varphi''$  of $A$ such that $ H(\varphi'')=H(\varphi') $.\\
proof: by the 1.1.
\\
2.3 Q.E.D.\\
\indent proof: By 2.1 and 2.2, for any execution $\varphi$ of $Z$ starting from any well-formed state $\sigma_z$,
there exists a sequential and terminating execution $\varphi''$ of $A$ and a history $h_c \in Compl(H(\varphi))$  such that $ h_c\sqsubseteq H(\varphi'') $.\\
3 Q.E.D.\\
By 1 and 2.
\end{proof}

\section{}
\begin{lemma}
 A concurrent object $Z$ is general  linearizable w.r.t its sequential specification if there exists an  ADT $A$, such that $Z$  is a concurrent  implementation  of $A$ w.r.t an abstraction function $AF$.
 \end{lemma}
\begin{proof}
1. For any concurrent execution $\pi$ of $Z$ starting from any well-formed state $\sigma_z$,
there exists a legal sequential execution $\pi'$ of $A$ starting from the state $AF(\sigma_z)$  and a history $h_c \in Compl(H(\pi))$  such that $ h_c\sqsubseteq H(\pi') $.
\\
proof: By the hypothesis, $Z$ is linearizable w.r.t. the ADT $A$.\\
2.  For any terminating execution  $\mu$: $ (\sigma_{a0},H_a,\sigma_{an})$ of $A$, there exists a sequential and terminating execution  $\mu'$:
 $ (\sigma_{z0}, H_{z},\sigma_{zn})$  of $Z$, such that $AF(\sigma_{z0})=\sigma_{a0} $, $AF(\sigma_{zn})=\sigma_{an}$ and  $H_a=H_z$.
 \\
\indent  2.1 Assume: $\mu=\sigma_{a0}\xrightarrow{\langle inv(op_1, n_1),ret(v_1) \rangle}
\sigma_{a1}\xrightarrow{\langle inv(op_2, n_2),ret(v_2) \rangle}
\sigma_{a2},\cdots, $  $\xrightarrow{\langle inv(op_n, n_n),ret(v_n) \rangle}$$\sigma_{an}$.
There exists $\sigma_{z0}$ and $\sigma_{z1}$ such that
$\sigma_{z0}\xrightarrow{inv(RF^{-1}(op_1),
n_1),ret(v_1)}\sigma_{z1}$.
\\
proof:
Since the abstraction function $AF$ is surjective, there exists a state of $Z$ is mapped
to $\sigma_{a0}$ by $AF$. Let $\sigma_{z0}$ be the state such that $AF(\sigma_{z0})= \sigma_{a0}$.
By  Definition 15,  there exists $\sigma_{z1}$ such that
$\sigma_{z0}\xrightarrow{inv(RF^{-1}(op_1),
n_1),ret(v_1)}\sigma_{z1}$ and $AF(\sigma_{z1})=\sigma_{a1}$.
\\
2.2 Q.E.D.\\
\indent \indent proof:
By 2.1 and Definition 15,
the execution $\mu'=\sigma_{z0}\xrightarrow{inv(RF^{-1}(op_1),
n_1),ret(v_1)}
\sigma_{z1}\xrightarrow{inv(RF^{-1}(op_2),
n_2),ret(v_2)}
\sigma_{z2},\cdots, $  $\xrightarrow{inv(RF^{-1}(op_n),
n_n),ret(v_n)}$$\sigma_{zn}$ is feasible.
Obviously, $H(\mu)=H({\mu}')$ and $AF(\sigma_{zi})= \sigma_{ai}$, for $i=1\cdots n$.
\\
3. Q.E.D.\\
 \indent proof:
 By 2, there exists a sequential and terminating execution $\pi''$  of $Z$, such that  $H(\pi'')=H(\pi')$. Since
  $ h_c\sqsubseteq H(\pi') $, $ h_c\sqsubseteq H(\pi'') $.
 \end{proof}

\textbf{Theorem 18.}\;\;
 A concurrent object $Z$ is strict linearizable  if there exists an  ADT $A$, such that $Z$  is a  concurrent   implementation  of $A$ w.r.t an injective abstraction function $AF$.
\\
\begin{proof}
1  For any terminating execution  $ (\sigma_{z},H_z,\sigma_{z}')$ of $Z$, there exists a sequential execution  $ (\sigma_{z},H_z',\sigma_{z}')$ of $Z$,   such that $ H_z\sqsubseteq H_z'$.
\\
1.1  For any terminating execution  $\mu$: $ (\sigma_{a0},H_a,\sigma_{an})$ of $A$, there exists a  sequential and terminating execution  $\mu'$:
 $ (\sigma_{z0}, H_{z},\sigma_{zn})$  of $Z$, such that $AF(\sigma_{z0})=\sigma_{a0} $, $AF(\sigma_{zn})=\sigma_{an}$ and  $H_a=H_z$.
 \\
\indent  1.1.1 Assume: $\mu=\sigma_{a0}\xrightarrow{\langle inv(op_1, n_1),ret(v_1) \rangle}
\sigma_{a1}\xrightarrow{\langle inv(op_2, n_2),ret(v_2) \rangle}
\sigma_{a2},\cdots, $  $\xrightarrow{\langle inv(op_n, n_n),ret(v_n) \rangle}$$\sigma_{an}$.
\\
There exists $\sigma_{z0}$ and $\sigma_{z1}$ such that
$\sigma_{z0}\xrightarrow{inv(RF^{-1}(op_1),
n_1),ret(v_1)}\sigma_{z1}$.
\\
proof:
Since the abstraction function $AF$ is injective, there exists a state of $Z$ is mapped
to $\sigma_{a0}$ by $AF$. Let $\sigma_{z0}$ be the state such that $AF(\sigma_{z0})= \sigma_{a0}$.
By  Definition 15,  there exists $\sigma_{z1}$ such that
$\sigma_{z0}\xrightarrow{inv(RF^{-1}(op_1),
n_1),ret(v_1)}\sigma_{z1}$ and $AF(\sigma_{z1})=\sigma_{a1}$.
\\
1.1.2 Q.E.D.\\
\indent \indent proof:
By 1.1 and Definition 15,
the execution $\mu'=\sigma_{z0}\xrightarrow{inv(RF^{-1}(op_1),
n_1),ret(v_1)}
\sigma_{z1}\xrightarrow{inv(RF^{-1}(op_2),
n_2),ret(v_2)}
\sigma_{z2},\cdots, $  $\xrightarrow{inv(RF^{-1}(op_n),
n_n),ret(v_n)}$$\sigma_{zn}$ is feasible.
Obviously, $H(\mu)=H({\mu}')$  and $AF(\sigma_{zi})= \sigma_{ai}$, for $i=1\cdots n$.
\\
1.2  For the terminating concurrent execution $ (\sigma_{z},H_z,\sigma_{z}')$ of $Z$, there exists a sequential execution $ (\sigma_{a}, H_{a},\sigma_{a}')$  of $A$, such that $AF(\sigma_{z})=\sigma_{a} $, $AF(\sigma_{z}')=\sigma_{a}'$ and  $H_z\sqsubseteq H_a$.
\\
proof: By Definition 15.
\\
1.3. Q.E.D.\\
By 1.1,
for the execution $ (\sigma_{a}, H_{a},\sigma_{a}')$  of $A$,
there exists a sequential execution  of $Z$:
 $ (\sigma_{zx}, H_z',\sigma_{zy})$ , such that $AF(\sigma_{zx})=\sigma_{a0} $, $AF(\sigma_{zy})=\sigma_{an}$ and  $H_a=H_z'$.
 Since $AF$ is injective, $\sigma_{zx}=\sigma_{z} $, $\sigma_{zy}=\sigma_{z}'$.
 Since  $H_z'= H_a$ and $H_z\sqsubseteq H_a$, $H_z\sqsubseteq H_z'$.
Thus,
for the concurrent execution of $Z$, $ (\sigma_{z},H_z,\sigma_{z}')$, there exists a sequential execution of $Z$, $ (\sigma_{z},H_z',\sigma_{z}')$   such that $ H_z\sqsubseteq H_z'$.
\\
2 $Z$ is general linearizable w.r.t its sequential specification.\\
proof: By Lemma 6.
\\
3 Q.E.D.\\
By 1 and 2, $Z$ is strict linearizable.
\end{proof}

\section{}

Fig. 6 shows the lock-free queue algorithm of Michael and Scott. The queue algorithm uses a linked list with Head and Tail pointers. The head pointer always points to the first node of the list.
The tail pointer points to the last node of the list in a quiescent state. The first node in the list acts as a dummy node o simplify certain list operations. The queue is meant to be empty when the list has only a dummy node. If the queue is not empty, the Dequeue method advances the head pointer nd returns the value of the new first node of the list, so the new first node becomes a new dummy node. If the queue is empty, then the Dequeue method returns EMPTY. The Enqueue method first appends a new node at the tail of the list, and later makes the tail pointer point to the new node. A thread cannot finish the Enqueue method in one atomic action, so other threads which observe that the tail pointer lags behind the end of the list will try to help the thread to advance the tail pointer before performing their own operations.
The concrete states of the algorithm are well-formed if the singly linked list does not contain cycles and the tail pointer points to the last node.

\begin{multicols}{2}
\multicolsep=3pt plus 2pt minus 2pt
\noindent
\tiny
class node\{\\
data\_t val;\\
node next;\}\\
class Queue \{\\
node Head,Tail;\\
void Enqueue(data\_t v);\\
data\_t Dequeue;()
 \}\\
 \\
void Enqueue(data\_t v) \{\\
local n, t, tn;\\
n := new\_node();\\
n.value :=v;\\
n.next=null;\\
while (true) \{\\
\indent t := Tail;\\
\indent tn := t.next;\\
\indent if (t = Tail) \{\\
\indent\indent if (tn = null) \{\\
 \indent\indent\indent if cas(\&(t.next),tn,n)\\
 \indent\indent\indent break; \}\\
\indent\indent else\\
\indent\indent cas(\&Tail, t, tn); \}\\
\indent\indent\}\\
\indent cas(\&Tail, t, n);\\
\}\\
\\
data\_t Dequeue() \{\\
 local h, t, hn, ret;\\
 while (true) \{\\
\indent h := Head;\\
\indent t := Tail;\\
\indent hn := h.next;\\
\indent if (h = Head)\\
 \indent if (h = t) \{\\
\indent \indent if (hn= null)\\
\indent\indent return EMPTY;\\
 \indent\indent cas(\&Tail, t, hn);\}\\
\indent else \{\\
\indent \indent ret := hn.value;\\
 \indent\indent if cas(\&Head,h,s);\\
\indent\indent return ret; \}\\
 \}\quad \}
\end{multicols}
\hrule
\begin{center} \tiny \textbf{Fig. 6.} the MS Lock-Free  Queue \end{center}

Consider a multiset data type with operations to add and remove elements from the multiset.
Let $\{x, \cdots  \}$ denote a multiset, one of whose member is $x$. mset represents the initial contents of the multiset. The notation $\varepsilon$ is different from the reserved value $EMPTY$ and indicates that a method does not return values.
The standard specification of a multiset is:
\[ Add(mset, e)=(mset \cup {e}, \varepsilon)\]
\[Remove(mset)=(mset', e),\]
 where  $mset=mset' \cup {e}$.\\
Consider an abstraction function which maps a concrete list pointed to by Head to the multiset consisting of the values of data fields of the list, and is formally defined as follows:
\[AF(Q)=\{ Q.Head.next.value \} \cup  AF(Q'),\]  where
$Q $ represents the MS lock-free queue and $Q'.Head = Q.Head.next$.
While the MS lock-free queue algorithm satisfies the multiset data type specification, this does not imply that  the algorithm is strict linearizable. This is because the abstraction function is not injective.

Consider a standard queue data type with Enqueue and Dequeue methods. The variable seq denotes an initial state of the atomic sequence,  $|seq|$ denotes the length of the  atomic  sequence.
The specification of the queue data type is defined as follows:
\[Enqueue(seq,x)=(seq^\smallfrown x, \varepsilon)\]
\[
Dequeue(seq)=
\begin{cases}
(seq',y), &  if \quad |seq|>0, \\
(seq, EMPTY),&  if \quad |seq|=0,
\end{cases}
\]
where $seq=y^\smallfrown seq'$.
The abstraction function maps a concrete list pointed to by Head to the value sequence of its data fields except the first data field, and is formally defined as follows:
\[
AF(Q)=
\begin{cases}
( ), \quad  if \quad Q.Head.next=null;  \\
Q.Head.next.value^\smallfrown AF(Q'),\; otherwise,
\end{cases}
\]
where $ Q'.Head=Q.Head.next$, and ( ) denotes an empty sequence. Under the abstraction, the data field of the first node is ignored by the users of the data structure.
Two lists which are the same except for the values of the data fields of their first nodes, are mapped to the same abstract value. Obviously, the abstraction function is not injective. Therefore, while the algorithm satisfies the queue data type specification, this does not imply the algorithm is  strict linearizable.

Consider a pseudo-queue data type, which is similar to a standard queue but does not allow dequeue operation when it contains one element. In practice, this may be because the first node must remain holding a global message.
The specification of the pseudo-queue data type is defined as follows:
\[Enqueue(seq,x)=(seq^\smallfrown x, \varepsilon)\]
\[
Dequeue(seq)=
\begin{cases}
(seq',y), &  if \quad |seq|>1,  \\
(seq, EMPTY),&  if \quad |seq|=1,
\end{cases}
\]
where $seq=y^\smallfrown seq'$.
Consider the following abstraction function:
\[
AF(Q)=
\begin{cases}
( ), \quad  if \quad Q.Head=null;  \\
(Q.Head.value)^\smallfrown AF(Q'),\; otherwise,
\end{cases}
\]
where $ Q'.Head=Q.Head.next$.
Note that the value of data filed of the first node is mapped to the first element of the pseudo-queue. Under the abstraction, the value of data filed of the first node is also observed by users. The abstraction function is injective.
Since we can show the algorithm satisfies the pseudo-queue specification, the MS lock-free queue algorithm is strict linearizable.

\end{document}